\documentclass[10pt]{article}
\RequirePackage{titlesec}
\RequirePackage{amsmath,esdiff}
\RequirePackage{amssymb}
\RequirePackage{authblk}

\RequirePackage[normalem]{ulem}
\RequirePackage{bbold,bm}
\RequirePackage{BOONDOX-cal}
\RequirePackage[colorlinks=true,linkcolor=blue,citecolor=red,urlcolor=red]{hyperref}
\RequirePackage{geometry}
\RequirePackage{setspace}
\RequirePackage{cite}
\RequirePackage{flushend}
\RequirePackage{mathrsfs}
\RequirePackage{hyperref}
\RequirePackage{graphicx}
\RequirePackage{cancel}
\RequirePackage{array,esint}
\RequirePackage[most]{tcolorbox}
\graphicspath{}
\titleformat{\section}{\fontsize{12}{12}\bfseries}{\thesection}{1em}{}
\begin{document}
\title{\bf Signatures of quantum geometry from exponential corrections to the black hole entropy}
\vskip 1cm
	\vskip 1cm
	\date{}
\author{\textbf{$\mathbf{Soham}$ $\mathbf{Sen}^{a\dagger}$, $\mathbf{Ashis}$ $\mathbf{Saha}^{a,b*}$ and $\mathbf{Sunandan}$ $\mathbf{Gangopadhyay}^{a\ddagger}$
\footnote{{\quad}\\
{${}^\dagger$ soham.sen@bose.res.in, sensohomhary@gmail.com}\\
{${}^*$sahaashis0007@gmail.com}\\
{${}^\ddagger$sunandan.gangopadhyay@bose.res.in, sunandan.gangopadhyay@gmail.com}}}}
\affil{{${}^a$ Department of Astrophysics and High Energy Physics}\\
{S.N. Bose National Centre for Basic Sciences}\\
{JD Block, Sector III, Salt Lake, Kolkata 700 106, India}\\
{${}^b$ Department of Physics}\\ {University of Kalyani, Kalyani 741235, India}}

\maketitle
\begin{abstract}
\noindent It has been recently shown in [\href{https://link.aps.org/doi/10.1103/PhysRevLett.125.041302}{Phys. Rev. Lett. 125 (2020) 041302}] that microstate counting carried out for quantum states residing on the horizon of a black hole leads to a correction of the form $\exp(-A/4l_p^2)$ in the Bekenstein-Hawking form of the black hole entropy. In this paper, we develop a novel approach to obtain the possible form of the spacetime geometry from the entropy of the black hole for a given horizon radius. The uniqueness of this solution for a given energy-momentum tensor has also been discussed. Remarkably, the black hole geometry reconstructed has striking similarities to that of noncommutative-inspired Schwarzschild black holes [\href{https://www.sciencedirect.com/science/article/pii/S0370269305016126}{Phys. Lett. B 632 (2006) 547}]. We also obtain the matter density functions using Einstein field equations for the geometries we reconstruct from the thermodynamics of black holes. These also have similarities to that of the matter density function of a noncommutative-inspired Schwarzschild black hole. The conformal structure of the metric is briefly discussed and the Penrose-Carter diagram is drawn. We then compute the Komar energy and the Smarr formula for the effective black hole geometry and compare it with that of the noncommutative-inspired Schwarzschild black hole. We also discuss some astrophysical implications of the solutions. Finally, we propose a set of quantum Einstein vacuum field equations, as a solution of which we obtain one of the spacetime solutions obtained in this work. We then show a direct connection between the quantum Einstein vacuum field equations and the first law of black hole thermodynamics.
\end{abstract}
\section{Introduction} 
The fact that gravity \cite{Einstein15,Einstein16} has a connection to thermodynamics is evident from our present insight that black holes and in particular their horizons behave like thermodynamical systems. It was demonstrated that isolated black hole horizons in equilibrium can be assigned a temperature $T=\frac{\hbar\kappa}{2\pi}$ where $\kappa$ is the surface gravity, and a thermodynamic entropy $S=\frac{A}{4l_p^2}$, where $l_p$ is the Planck length. With these identifications, the evolution of the horizon from one equilibrium state to another obeys a relation identical to the first law of thermodynamics. 
 It is to be noted however that the ``\textit{area by four law}" for the black hole entropy (known as Bekenstein-Hawking black hole entropy \cite{Bekenstein,Bekenstein2,Hawking,Hawking2,Hawking3,Hawking4}) and the temperature of the horizon (known as the Hawking temperature \cite{Hawking,Hawking2,Hawking3}) comes from a semi-classical treatment and not from a theory of quantum gravity which is a unification of gravity with quantum mechanics. The quantum gravity programme would involve a microstate counting of states on the black hole horizon. In this process one identifies local horizon microstates of the black hole and can derive an area spectrum. In this approach the main role is played by the quantum representation of the black hole horizon.
 
\noindent In order to unify gravity with quantum mechanics, there have been several attempts in the last few decades. String theory \cite{ACV,KOPAPR}, loop quantum gravity \cite{ROVELLI,CARLIP} and noncommutative geometry \cite{Girelli} are such quantum gravity theories. An important thrust area where these theories have been applied is in trying to understand the microscopic origin of black hole entropy and also look for quantum corrections to the semi-classical ``\textit{area by four}" law for the entropy of the black hole. 
 In string theory and loop quantum gravity, microstate counting of black hole states has been carried out explicitly to derive the entropy of a black hole. This process not only yields the $\frac{A}{4l_p^2}$ term but also generates corrections to it. These corrections are expressed as an expansion of the $\frac{l_p^2}{A}$ terms, logarithmic correction \cite{Strominger, RD, Ashtekar,ParthaKaul, Lewandowski,Meissner,Ghosh2,MandalSen,RBSGSKM,Holkar,SGDKR,SG0} and an exponential correction \cite{Holkar2,Ghosh}. The modified entropy structure, in general, can be written as
\begin{equation}\label{0.2}
	S=\frac{A}{4l_p^2}+\beta_0\ln\left[\frac{A}{4l_p^2}\right]+\frac{4\beta_1l_p^2}{A}+\cdots+\exp\left(-\alpha_0\frac{A}{4l_p^2}\right)+\cdots
\end{equation}
where $\beta_0,\beta_1,\cdots,\alpha_0$ are constants. These corrections are in decreasing order of importance for large horizon areas compared to $l_p^2$. However, for small areas of the black horizon, the exponential correction would be the dominating correction term. Recently exponential corrections to the area law for entropy has been obtained in \cite{Ghosh} by using the horizon geometry approach. It comes from a microstate counting of horizon states with an equidistant area spectrum and reads
\begin{equation}\label{0.3}
S=\frac{A\ln 2}{8\pi\gamma l_p^2}+\exp\left[-\frac{A\ln 2}{8\pi\gamma l_p^2}\right]~.
\end{equation}
Choosing the parameter $\gamma$ to be $\frac{\ln 2}{2\pi}$, the above expression for entropy reduces to the following form \cite{Ghosh}
\begin{equation}\label{0.3A}
	S=\frac{A}{4l_p^2}+\exp\left[-\frac{A}{4l_p^2}\right]~.
\end{equation} 
It can be observed from eq.(\ref{0.3A}) that the leading order term resembles the celebrated Bekenstein-Hawking black hole entropy and the sub-leading term gives exponentially suppressed corrections to the classical result. A discrete area spectrum from which eq.(\ref{0.3A}) follows  is inevitably related to a non-trivial commutator between geometrical observables (noncommutativity), as in loop quantum gravity \cite{Rovelli}, and in noncommutative geometry \cite{Doplicher,Doplicher2,Douglas}. Remarkably, the above structure of the entropy has a striking resemblance to the entropy of a noncommutative inspired black hole \cite{Nicolini, Nicolini2,Chaichian, Nicolini3,SGRB}. 
Such a geometry arises from ordinary commutative Einstein field equations sourced by matter that has Gaussian density profile, with the width of the Gaussian profile being of the order of Planck length. In particular, point-like objects are replaced by smeared structures with the important feature of there being no curvature singularity of the metric at the origin. In \cite{RBSGSKM}, it was demonstrated that such a Gaussian density profile can be derived from the Voros star product \cite{Voros} between spatial coordinates, which in turn relates this black hole geometry with noncommutativity.
This motivates us to explore whether one can derive an effective black hole geometry from the above entropy structure and compare it with that coming from a noncommutative inspired black hole. We would like to stress that deriving a black hole geometry from the entropy of a black hole has not been reported earlier in the literature. Our work is therefore the first in this direction. The implication of this work would be very interesting as this would possibly give a way of identifying the underlying matter density function of a black hole whose entropy has exponential corrections, which in this case would be similar to matter density function for the noncommutative geometry. Our work would also reinforce the connection between thermodynamics and gravity.

\noindent We would now like to recall the metric structure of a noncommutative inspired Schwarzschild black hole (NS black hole) \cite{Nicolini, Nicolini2,Chaichian, Nicolini3,SGRB} 
\begin{equation}\label{0.5}
	f_\theta(r)=1-\frac{2 M_\theta(r)}{r}
\end{equation}
where the form of $M_\theta(r)$ is given by
\begin{equation}\label{0.6}
	M_\theta(r)=\frac{2M}{\sqrt{\pi}}\gamma\left[\frac{3}{2},\frac{r^2}{4\theta}\right]
\end{equation}
with $\theta=l_p^2$ being the noncommutative parameter, $l_p$ being the Planck length, $M$ being the mass of the black hole, and $\gamma$ being the lower incomplete gamma function which is given by \cite{Gradshteyn}
\begin{equation}\label{0.7}
	\gamma\left[\frac{3}{2},\frac{r^2}{4\theta}\right]=\int_0^{\frac{r^2}{4\theta}}dy~y^{\frac{1}{2}}e^{-y}~.
\end{equation}
An important feature of this metric structure (eq.(\ref{0.5})) is that in the limit $r\rightarrow 0$, the lapse function $f(r)$ is divergence free ($\lim\limits_{r\rightarrow0}f_\theta(r)=1$) showing the avoidance of the physical singularity of a Schwarzschild black hole at $r=0$. 

\noindent The metric structure (eq.(\ref{0.5})) has a de Sitter core in the $r\rightarrow 0$ limit. The Penrose-Carter diagram for the noncommutative inspired Schwarzschild black hole can be found in \cite{Arraut}. In the regime $\frac{r^2}{4\theta}\gg 1$, the metric reads
\begin{equation}\label{0.7a}
f_{\theta}(r)=1-\frac{2M}{r}+\frac{2M}{\sqrt{\pi\theta}}\left(1+\frac{2\theta}{r^2}\right)e^{-\frac{r^2}{4\theta}}+\mathcal{O}(\theta^{\frac{3}{2}}e^{-\frac{r^2}{4\theta}})
\end{equation}
and the Bekenstein-Hawking entropy upto $\mathcal{O}\left(\sqrt{\theta}e^{-\frac{M^2}{\theta}}\right)$ reads \cite{SGRB}
\begin{equation}\label{0.8}
	S=\frac{4\pi M^2}{\theta}-16\sqrt{\frac{\pi}{\theta^3}}M^3\left(1+\frac{\theta}{M^2}\right)e^{-\frac{M^2}{\theta}}
\end{equation}
where we have restored the Planck length in the first term. The above result can be interpreted in a very nice way. The first term can be written as $\frac{A}{4 l_p^2}$ with $A=16\pi M^2$ being the area of the standard Schwarzschild black hole. The second term is precisely the exponential correction to the area law of the Schwarzschild black hole. It is therefore very crucial to observe that exponential corrections do arise in the black hole entropy from noncommutative inspired black hole geometries. The above result can be recast as
\begin{equation}\label{0.9}
S=S_{Sch.}-\frac{2\theta S_{Sch.}^{\frac{3}{2}}}{\pi}\left(1+4\pi S_{Sch.}^{-1}\right)e^{-\frac{S_{Sch.}}{4\pi}}
\end{equation}
where $S_{Sch.}$ is the entropy of the standard Schwarzschild black hole.

\noindent In this paper we have exploited the modified entropy relation eq.(\ref{0.3}) to obtain the possible metric structure of the underlying black hole geometry. To begin with, we formulate a simple technique to extract the metric structure from the Bekenstein-Hawking entropy formula for a Reissner-Nordstr\"{o}m (RN) black hole spacetime. Using this technique, we have then obtained the metric structure from the modified entropy formula (including the exponential correction term) given in eq.(\ref{0.3}). 

\vskip 0.08cm

\section{Prescription to compute the metric} 
The general structure of first law of black hole thermodynamics reads \cite{Bekenstein2,Hawking2,Hawking4}
\begin{equation}\label{1.0}
	dM=TdS + \phi dQ+ \Omega dJ
\end{equation}
\noindent where $dM$ is the change in the mass of the black hole, $TdS$ represents the Hawking temperature-black hole entropy part, $\phi dQ$ represents the electrostatic potential and charge part, and $\Omega dJ$ represents the angular momentum part. Now one can write down the Bekenstein-Hawking black hole entropy in terms of the event horizon radius $r_+$ as $S=\pi r_+^2$.
From the first law of black hole thermodynamics (given in eq.(\ref{1.0})), it is possible to define the Hawking temperature in the following way
\begin{equation}\label{1.2}
	\left(\frac{\partial S}{\partial M}\right)_{Q,J}=\frac{1}{T}~.
\end{equation}
On the other hand, the Hawking temperature in eq.(\ref{1.2}) is related to the surface gravity ($\kappa$) as follows \footnote{Note that we are considering black holes where the line element is given as $ds^2=-f(r)dt^2+\frac{1}{f(r)}dr^2+r^2d\Omega^2$.}
\begin{equation}\label{1.3}
	T=\frac{\kappa}{2\pi}=\frac{f'(r_+)}{4\pi}~.
\end{equation} 
The thermodynamic relation (eq.(\ref{1.2})) and the geometric relation (given in eq.(\ref{1.3})) are the fundamental inputs for our prescription. The only other input our prescription requires is the structure of the event horizon radius ($r_+$) in terms of the black hole parameters. The uniqueness of our solution comes from the Einstein's field equations.\\
We now briefly demonstrate our prescription for a simple case, namely, the static spherically symmetric non-extremal  Reissner-Nordstr\"{o}m black hole solution.
The event horizon radius for this solution is known to have the following expression $r_{\pm}=M\pm\sqrt{M^2-Q^2}$, where the $\pm$ sign represents the outer and inner event horizons. The Bekenstein-Hawking entropy of the black hole now reads
\begin{equation}\label{1.5}
	S=\pi r_+^2=\pi(2M^2-Q^2+2M\sqrt{M^2-Q^2})~.
\end{equation}
\noindent Using the relation given in eq.(\ref{1.2}) and under the condition $Q\ll r_+$, 
and using the geometric definition of Hawking temperature (given in eq.(\ref{1.3})) to obtain the first derivative (with respect to the radial coordinate) of the lapse function at the horizon, we have $f'(r_+)=\frac{1}{r_+}-\frac{Q^2}{r_+^3}$~.
We now introduce two physical conditions which the black hole metric must satisfy. First,
	the black hole metric must produce the flat spacetime metric at the infinite distance. This means the lapse function must obey the condition $\lim\limits_{r\rightarrow\infty}f(r)=1$,
	and second, the lapse function $f(r)$ must vanish at the horizon, that is, we must have $f(r_+)=0$.
Now keeping in mind the form of $f'(r_+)$ (given earlier), we take the following ansatz for the form of the metric
\begin{equation}\label{1.9}
	f(r)=\sum\limits_{k=0}^\infty a_kr^{-k}
\end{equation}
where the coefficients $a_k$ has to be determined from the above physical conditions. 
The first physical condition readily gives $\lim\limits_{r\rightarrow\infty}f(r)=a_0=1$.
Eq.(\ref{1.9}) now reduces to the following structure $f(r)=1+\sum_{k=1}^\infty a_kr^{-k}~.$
Computing $f'(r_+)$ from the above equation and comparing it with $f'(r_+)$ obtained earlier yields the condition
$\frac{(a_2-Q^2)}{r_+^3}+\frac{2a_3}{r_+^4}+\cdots=0~.$
This implies
$a_2=Q^2, 	a_k=0~,~k=3,4,\cdots~$.
Using these values along with $a_0=1$, the metric form in eq.(\ref{1.9}) can be recast as $f(r)=1+\frac{a_1}{r}+\frac{Q^2}{r^2}~.$
The unknown coefficient $a_1$ can be obtained from the second physical condition, which gives $a_1=-\left(r_++\frac{Q^2}{r_+}\right)=-2M~$.
In obtaining this, we have used the form of $r_+$ in terms of the black hole parameters.
Using the form of $a_1$, the sought metric takes the form
\begin{equation}\label{1.20}
	f(r)=1-\frac{2M}{r}+\frac{Q^2}{r^2}~.
\end{equation}
We now discuss the uniqueness of our procedure. It should be noted that constructing a function of two variables $f(r,r_+)$ from the conditions $f(r+r_+,r_+)=0$ and $f(r=\infty,r_+)=1$ and the value of its first partial derivative $f'(r_+)$ is not unique as can be seen easily, for example, by adding a function $g(r,r_+)=(r-r_+)^2\exp(-a(r-r_+))$ to $f(r,r_+)$ with $f+g$ having the same property as $f$. To fix the uniqueness of $f(r,r_+)$ obtained from our approach, we use the Einstein's field equations with an appropriate energy-momentum tensor. In the Maxwell case, the form of the energy momentum tensor fixes the field equation to be
\begin{equation}\label{1.20a}
\frac{d}{dr}(rf(r))=1-\frac{Q^2}{r^2}~.
\end{equation}
Computing the left hand side of the above equation with eq.(\ref{1.20}) yields the right hand side thereby showing the uniqueness of the solution. The alternative way of understanding this is to consider a modified lapse function of the form $h(r)=f(r)+g(r)$, where $h(r)$ has the required properties of $f(r)$. The field equation (eq.(\ref{1.20a})) gives $g(r)=\frac{\mathcal{C}_1}{r}$, where $\mathcal{C}_1$ is an integration constant. Hence, $h(r)$ has the form, $h(r)=1+\frac{\mathcal{C}_1-2M}{r}+\frac{Q^2}{r^2}$. Now from the second condition,  $h(r)$ must vanish at the horizon radius for the RN black hole. Using this condition, we finally obtain the value of constant $\mathcal{C}_1$ to be zero showing the uniqueness of the solution obtained. The take away lesson is that the above procedure together with the explicit form of horizon radius and the consistency condition from the Einstein's field equations gives the metric structure from the Bekenstein-Hawking entropy formula. Note that the procedure does not solve the field equations of gravity, the solution comes from thermodynamics and the uniqueness of the solution gets fixed from the field equation. This also establishes the subtle connection between thermodynamics and gravity \cite{TJacobson}.    

\vskip 0.08cm

\section{Metric structure from the modified Bekenstein-Hawking black hole entropy}
We would like to mention before starting that a careful calculation of the entropy from the microstate of the black hole in \cite{Ghosh} indicates that there should be an area term in front of the exponential correction in eq.(\ref{0.3}). This can be seen by looking at the number of microstates on the horizon \cite{Ghosh} given by $\Omega=\frac{(\sum_i s_i)!}{\prod_is_i!}$, where $\sum_i s_i$ is the total number of tiles in the tessellation. A detailed analysis now gives the Bekenstein-Hawking entropy of the black hole to be
\begin{equation}\label{1.21}
S=\frac{k_BA}{4l_p^2}-\frac{k_BA}{4l_p^2\ln2}e^{-\frac{A}{4l_p^2}}~.
\end{equation}
 This makes us consider the following general structure of the Bekenstein-Hawking black hole entropy
\begin{equation}\label{1.22}
	S=\frac{k_BA}{4l_p^2}- k_B\left(\frac{A}{l_p^2}\right)^ne^{-\frac{A}{4l_p^2}}
\end{equation}  
where $n=\{\cdots,-\frac{1}{2},0,\frac{1}{2},1,\frac{3}{2},\cdots\}$. Hence, we shall consider $n=0,\frac{1}{2},1$. To proceed further we start with the general structure for a static spherically symmetric black hole \cite{Arbey}
\begin{equation}\label{1.23}
ds^2=-f(r)c^2dt^2+\frac{1}{f(r)}dr^2+h(r)d\Omega^2
\end{equation}
where $d\Omega^2=d\theta^2+\sin^2\theta~ d\phi^2$. Generally $f(r)$ is known as the lapse function of the black hole and for standard spherically symmetric black holes $h(r)=r^2$. For a generalized analysis, we shall continue with the line element given in eq.(\ref{1.23}). Our initial demand is that the entropy in eq.(\ref{1.22}) can be written as $S=\frac{k_B\tilde{A}}{4l_p^2}$. The event horizon area of the black hole (whose line element is given by eq.(\ref{1.23})) reads
\begin{equation}\label{1.24}
\tilde{A}=\int_0^\pi d\theta\int_0^{2\pi}d\phi \sqrt{g_{\theta\theta}g_{\phi\phi}}=4\pi h(r_+)~.
\end{equation}
From the above equation, it is straightforward to infer that the modified event horizon radius of the black hole is $\tilde{r}_+=\sqrt{h(r_+)}$.

\noindent Hence, the entropy of the black hole in terms of $h(r_+)$ is given by 
\begin{equation}\label{1.24a}
\begin{split}
S&=\frac{k_B\tilde{A}}{4l_p^2}\\&=\frac{k_B\pi h(r_+)}{l_p^2}
\end{split}
\end{equation}
where $r_+=\frac{2GM}{c^2}$. Comparing the entropy in the above equation in terms of $\tilde{A}$ with the form of the entropy in eq.(\ref{1.22}) (for $n=0$), we obtain
\begin{equation}
\begin{split}
h(r_+)&=r_+^2-\frac{l_p^2}{\pi}e ^{-\frac{\pi r_+^2}{l_p^2}}~. 
\end{split}
\end{equation}
From the above equation, we get the form of the event horizon radius of this modified black hole in terms of the Schwarzschild radius $r_+$ (for $n=0$) as
\begin{equation}\label{1.25}
	\tilde{r}_+=\sqrt{h(r_+)}\simeq{r_+}\left(1-\frac{l_p^2}{2\pi r_+^2}e^{-\frac{\pi r_+^2}{l_p^2}}\right)
\end{equation}
since $e^{-\frac{A}{4l_p^2}}\ll 1$.
We shall now follow our prescription. Substituting the form of the Schwarzschild radius $r_+$ in terms of the mass of the black hole in eq.(\ref{1.22}) and calculating $\frac{\partial S}{\partial M}$, we obtain
\begin{equation}\label{1.25a}
\begin{split}
\frac{\partial S}{\partial M}&=\frac{1}{T}=\frac{4\pi k_B r_+G}{l_p^2 c^4}\left(1+e^{-\frac{\pi r_+^2}{l_p^2}}\right)\\
\implies T&\simeq  \frac{l_p^2c^4}{4\pi k_BGr_+}\left(1- e^{-\frac{\pi r_+^2}{l_p^2}}\right)
\end{split}
\end{equation}
where $T$ is the Hawking temperature of the black hole.
This fixes the form of $f'(\tilde{r}_+(r_+))$ to be 
\begin{equation}\label{1.27}
	f'(\tilde{r}_+(r_+))\cong\frac{1}{r_+}\left(1- e^{-\frac{\pi r_+^2}{l_p^2}}\right)~.
\end{equation}
Proceeding as before, we take the following ansatz for the form of the metric
\begin{equation}\label{1.28}
	f(r)=\sum\limits_{k=0}^\infty a_kr^{-k}+\sum\limits_{n=-\infty}^\infty b_kr^ke^{-\frac{\pi r^2}{l_p^2}}~.
\end{equation}
From eq.(\ref{1.28}), we observe that in the $l_p^2\rightarrow 0$ limit, the metric structure has the same form as the initial ansatz in eq.(\ref{1.9}). From eq.(\ref{1.25}), we observe that in this limit, $\tilde{r}_+\rightarrow r_+$. 
Therefore, using the condition (2), enumerated in the text, we can obtain the values of the non-vanishing `$a_k$' coefficients in the following way. We first note that in the $l_p^2\rightarrow 0$ limit
\begin{equation}\label{1.29}
	f(r_+)=a_0+\frac{a_1}{r_+}+\frac{a_2}{r_+^2}+\cdots=0~.
\end{equation}
Now making use of the asymptotic condition for $f(r)$, we obtain $a_0=1$ which from eq.(\ref{1.29}) gives us the condition 
\begin{equation}\label{1.29a}
\frac{a_1}{r_+}+\frac{a_2}{r_+^2}+\cdots=-1~.
\end{equation}
From eq.(\ref{1.27}), we observe that in this limit, $f'(\tilde{r}_+)$ reduces to the following form 
\begin{equation}\label{1.30}
	f'(r_+)=\frac{1}{r_+}~.
\end{equation}
In this same $l_p^2\rightarrow 0$ limit, using the procedure similar to the RN black hole case, we obtain from eq.(\ref{1.28})
\begin{equation}\label{1.31}
\begin{split}
f'(r_+)&=-\frac{a_1}{r_+^2}-\frac{2a_2}{r_+^2}-\cdots\\
&=-\frac{1}{r_+}\left(\frac{a_1}{r_+}+\frac{a_2}{r_+^2}+\frac{a_3}{r_+^3}\cdots\right)-\frac{a_2}{r_+^2}-\frac{2a_3}{r_+^2}+\cdots\\
&=\frac{1}{r_+}-\frac{a_2}{r_+^2}-\frac{2a_3}{r_+^2}+\cdots~.
\end{split}
\end{equation}
Comparing the above result with eq.(\ref{1.30}), we obtain $a_2=a_3=\cdots=0$. The Schwarzschild limit in the $l_p^2\rightarrow 0$ case gives $a_1=r_+=-\frac{2GM}{c^2}$. 
Eq.(\ref{1.28}) then simplifies to
\begin{equation}\label{1.33}
	f(r)=1-\frac{r_+}{r}+\sum\limits_{k=-\infty}^\infty b_kr^ke^{-\frac{\pi r^2}{l_p^2}}~.
\end{equation}
To obtain the unknown coefficients in the second term, we will use the condition $f(\tilde{r}_+(r_+))=0$ and compare the forms of $f'(\tilde{r}_+(r_+))$ obtained from the Hawking temperature and from the metric ansatz in eq.(\ref{1.28}). As $e^{-\frac{\pi r_+^2}{l_p^2}}\neq 0$, this leads to the condition
\begin{equation}\label{1.34}
	\begin{split}
		f(\tilde{r}_+)&=1-\frac{r_+}{\tilde{r}_+}+\left(\cdots+\frac{b_{-1}}{\tilde{r}_+}+b_0+b_1\tilde{r}_++\cdots\right)e^{-\frac{\pi \tilde{r}_+^2}{l_p^2}}\\
		&=\left(\cdots+\frac{- l_p^2+2\pi b_{-2}}{2\pi r_+^2}+\frac{b_{-1}}{r_+}+b_0+\cdots\right)e^{-\frac{\pi r_+^2}{l_p^2}}\\&=0~.
	\end{split}
\end{equation}
As $e^{-\frac{\pi r_+^2}{l_p^2}}\neq 0$, we get from eq.(\ref{1.34}) the following relation
\begin{equation}\label{1.35a}
	\cdots+\frac{b_{-1}}{r_+}+b_0+b_1 r_++\cdots=\sum\limits_{k=-\infty}^{\infty}b_kr_+^k=\frac{l_p^2}{2\pi r_+^2}~.
\end{equation}
The form of $f'(\tilde{r}_+(r_+))$ can be obtained from eq.(\ref{1.33}) to be
\begin{equation}\label{1.36}
\begin{split}
		f'(\tilde{r}_+(r_+))=&\frac{r_+}{\tilde{r}_+^2}+\biggr(\cdots-\frac{b_{-1}}{\tilde{r}_+^2}+b_1+2b_2\tilde{r}_+\cdots\biggr)e^{-\frac{\pi \tilde{r}_+^2}{l_p^2}}-\frac{2\pi \tilde{r}_+}{l_p^2}\left(\cdots+\frac{b_{-1}}{\tilde{r}_+}+b_0+b_1\tilde{r}_++\cdots\right)e^{-\frac{\pi \tilde{r}_+^2}{l_p^2}}~.
	\end{split}
\end{equation}
Substituting the form of the perturbed horizon radius from eq.(\ref{1.25}) in eq.(\ref{1.36}), we obtain
\begin{equation}\label{1.37a}
	\begin{split}
		f'(\tilde{r}_+)&\cong\frac{1}{r_+}+\frac{l_p^2}{\pi r_+^3}e^{-\frac{\pi r_+^2}{l_p^2}}+\biggr[\cdots-\frac{3b_{-2}}{r_+^3}-\frac{2b_{-1}}{r_+^2}+b_2 r_+\cdots+\frac{1}{r_+}\left(\sum\limits_{k=-\infty}^{\infty}b_kr_+^k\right)-\frac{b_0}{r_+}\biggr]e^{-\frac{\pi r_+^2}{l_p^2}}\\&-\frac{2\pi r_+}{l_p^2}\left(\sum\limits_{k=-\infty}^{\infty}b_kr_+^k\right)e^{-\frac{\pi r_+^2}{l_p^2}}~.
	\end{split}
\end{equation}
Using the form of $\sum\limits_{k=-\infty}^{\infty}b_kr_+^k$ from eq.(\ref{1.35a}) in eq.(\ref{1.37a}), we can recast $f'(\tilde{r}_+)$ in the following form
\begin{equation}\label{1.37}
	\begin{split}
		f'(\tilde{r}_+(r_+))=&\frac{1}{r_+}+\cdots-\frac{4b_{-3}e^{-\frac{\pi r_+^2}{l_p^2}}}{r_+^4}-\left[\frac{3\alpha l_p^2}{2\pi }+3b_{-2}\right]\frac{e^{-\frac{\pi r_+^2}{l_p^2}}}{r_+^3}
		-\frac{2b_{-1}}{r_+^2}e^{-\frac{\pi r_+^2}{l_p^2}}+b_2r_+e^{-\frac{\pi r_+^2}{l_p^2}}+\cdots+\frac{\alpha-b_0}{r_+}e^{-\frac{\pi r_+^2}{l_p^2}}~.
\end{split}
\end{equation}
We now equate the above form of $f'(\tilde{r}_+(r_+))$ with the form of $f'(\tilde{r}_+(r_+))$ obtained earlier in eq.(\ref{1.27}) from the first law of black hole thermodynamics.
This leads to the form of the unknown coefficients in eq.(\ref{1.28}) as $b_{-2}=\frac{l_p^2}{2\pi}$ and	$\cdots=b_{-3}=b_{-1}=b_0=b_1=b_2=\cdots=0~.$
Hence, using the values of the above coefficients, the final form of $f(r)$ reads
\begin{equation}\label{1.42}
	f(r)=1-\frac{2GM}{rc^2}+\frac{l_p^2}{2\pi r^2}e^{-\frac{\pi r^2}{l_p^2}}~.
\end{equation}
As a consistency check, one can obtain the form of  $f'(r)$ from the above equation to be $f'(r)=\frac{r_+}{r^2}-\frac{l_p^2}{\pi r^3}e^{-\frac{\pi r^2}{l_p^2}}-\frac{1}{r}e^{-\frac{\pi r^2}{l_p^2}}$~. Substituting $\tilde{r}_+$ in $f'(r)$, we obtain
\begin{equation}\label{1.421}
\begin{split}
f'(\tilde{r}_+)&=\frac{r_+}{\tilde{r}_+^2}-\frac{l_p^2}{\pi\tilde{r}_+^3}e^{-\frac{\pi\tilde{r}_+^2}{l_p^2}}-\frac{1}{\tilde{r}_+}e^{-\frac{\pi\tilde{r}_+^2}{l_p^2}}\\
&\simeq\frac{r_+}{r_+^2\left(1-\frac{l_p^2}{\pi r_+^2}e^{-\frac{\pi r_+^2}{l_p^2}}\right)}-\frac{l_p^2}{\pi r_+^3}e^{-\frac{\pi r_+^2}{l_p^2}}-\frac{1}{r_+}e^{-\frac{\pi r_+^2}{l_p^2}}\\
&\simeq \frac{1}{r_+}+\frac{l_p^2}{\pi r_+^3}e^{-\frac{\pi r_+^2}{l_p^2}}-\frac{l_p^2}{\pi r_+^3}e^{-\frac{\pi r_+^2}{l_p^2}}-\frac{1}{r_+}e^{-\frac{\pi r_+^2}{l_p^2}}\\
&=\frac{1}{r_+}\left(1-e^{-\frac{\pi r_+^2}{l_p^2}}\right)~.
\end{split}
\end{equation}
This is same as the form of $f'(\tilde{r_+})$ obtained in eq.(\ref{1.27}), indicating the same Hawking temperature as in eq.(\ref{1.25a}). Throughout the calculation it is important to keep in mind that any higher orders of the factor $e^{-\frac{\pi r_+^2}{l_p^2}}$ has been neglected.

\noindent Our next aim is to obtain the form of $h(r)$. We already know the form of $h(r_+)$ from eq.(\ref{1.24a}). Since
 in the asymptotic limit $r\rightarrow\infty$, $h(r)\rightarrow r^2$, we know that the leading order term in $h(r)$ should be $r^2$. We can therefore straight away write down the structure of $h(r)$ to be
\begin{equation}\label{1.422}
h(r)=r^2-\frac{l_p^2}{\pi}e^{-\frac{\pi r^2}{l_p^2}}~.
\end{equation}
The non-vanishing components of the Einstein tensor for the line element of the black hole given in eq.(\ref{1.23}) reads
\begin{align}
G_{tt}&=f(r)\left(\frac{f(r)h^{'2}(r)}{4h^2(r)}-\frac{f(r)h''(r)}{h(r)}+\frac{1}{h(r)}-\frac{f'(r)h'(r)}{2h(r)}\right)~,\label{Ein00}\\
G_{rr}&=\frac{1}{f(r)}\left(\frac{f(r)h^{'2}(r)}{4h^2(r)}-\frac{1}{h(r)}+\frac{f'(r)h'(r)}{2h(r)}\right)~,\label{Ein11}\\
G_{\theta\theta}&=\frac{1}{4}\left(2f'(r)h'(r)+2f(r)h''(r)+2h(r)f''(r)-\frac{f(r)h^{'2}(r)}{h(r)}\right)~,\label{Ein22}\\
G_{\phi\phi}&=\frac{\sin^2\theta}{4}\left(2f'(r)h'(r)+2f(r)h''(r)+2h(r)f''(r)-\frac{f(r)h^{'2}(r)}{h(r)}\right)~.\label{Ein33}
\end{align}
Now in the Einstein equations $G_{\mu\nu}=\frac{8\pi G}{c^4}T_{\mu\nu}$, we make use of the non-vanishing components of the Einstein tensor eq.(\ref{Ein00}-\ref{Ein33}) and substituting the forms of $h(r)$ and $f(r)$, we obtain the forms of the non-vanishining components of the energy momentum tensor to be
\begin{align}
T^{t}_{~t}=&-\frac{c^4}{8\pi G r^2}\left(\left(1-\frac{r_+}{r}\right)
\left(1+\frac{l_p^2}{\pi r^2}+\frac{4\pi r^2}{l_p^2}\right)+\frac{l_p^2}{2\pi r^2}\right)e^{-\frac{\pi r^2}{l_p^2}}\nonumber\\
\simeq& -\frac{c^4}{8\pi G r^2}\frac{4\pi r^2}{l_p^2}\left(1-\frac{r_+}{r}\right)e^{-\frac{\pi r^2}{l_p^2}}=-\frac{c^4}{2Gl_p^2}\left(1-\frac{r_+}{r}\right)e^{-\frac{\pi r^2}{l_p^2}}\label{EM00}~,\\
T^r_{~r}=&\frac{c^4}{8\pi Gr^2}\left(\left(1-\frac{r_+}{r}\right)\left(1+\frac{l_p^2}{\pi r^2}\right)-\frac{l_p^2}{2\pi r^2}\right)e^{-\frac{\pi r^2}{l_p^2}}\simeq \frac{c^4}{8\pi Gr^2}\left(1-\frac{r_+}{r}\right)e^{-\frac{\pi r^2}{l_p^2}}\label{EM11},\\
T^\theta_{~\theta}=&-\frac{c^4}{8\pi G r^2}\left(\frac{1}{2}\left(1-\frac{r_+}{r}\right)\left(1+\frac{l_p^2}{\pi r^2}+\frac{2\pi r^2}{l_p^2}\right)+\frac{3r_+}{2r}\left(1+\frac{l_p^2}{\pi r^2}+\frac{2\pi r^2}{3 l_p^2}\right)\right)e^{-\frac{\pi r^2}{l_p^2}}\nonumber\\\simeq& -\frac{c^4}{8Gl_p^2}\left(1-\frac{r_+}{r}\right)e^{-\frac{\pi r^2}{l_p^2}}\label{EM22},\\
T^{\phi}_{~\phi}\simeq&-\frac{c^4}{8Gl_p^2}\left(1-\frac{r_+}{r}\right)e^{-\frac{\pi r^2}{l_p^2}}~.
\label{EM33}
\end{align}
It is important to note that in our analysis in which the usual classical general relativistic equations are valid, there cannot be any vacuum black hole solutions corresponding to the field equations. In order to obtain vacuum solutions, one needs to modify the Einstein field equations which would then lead to quantum Einstein equations as discussed in section (\ref{QEin}).
For the special case $h(r)=r^2$, substituting the form of $f(r)$ from eq.(\ref{1.42}) in the Einstein's field equation, $G_{\theta\theta}=8\pi T_{\theta\theta}$, we obtain the general form of the matter density function as follows
\begin{equation}\label{1.42a}
\rho(r)=\frac{c^4}{8\pi G r^2}\left(1+\frac{l_p^2}{2\pi r^2}\right)e^{-\frac{\pi r^2}{l_p^2}}\simeq \frac{c^4}{8\pi Gr^2}e^{-\frac{\pi r^2}{l_p^2}}~.
\end{equation}
This is an important finding in our paper since following our approach, we not only obtain the lapse function from the modified Bekenstein-Hawking black hole entropy but also get the form of the matter density needed to obtain such a metric structure. It is again important to observe that when $r\rightarrow 0$, then the matter density in eq.(\ref{1.42a}) diverges. It is important to note that the modified entropy relation in eq.(\ref{1.22}) is a result derived from a loop quantum gravity analysis. On the other hand loop quantum gravity does indicate towards the existence of a fundamental minimal length in nature which is of the order of the Planck length, $l_p$. Hence, if we want to probe the results obtained in our analysis below the Planck length scale, it will lead to some inconsistencies as can be observed from probing the matter density in the $r\rightarrow 0$ limit. Now black hole must have singularities at the $r=0$ point. In a quantum gravity analysis with the existence of an intrinsic fundamental length scale, we cannot really go below $r<l_p$. Hence, what is really a black hole from a quantum gravity perspective is a matter of further analysis.
Following a similar procedure, we can now obtain the metric structures for different $n$ values in eq.(\ref{1.22}).

\begin{figure}[ht!]
\begin{center}
\includegraphics[scale=0.36]{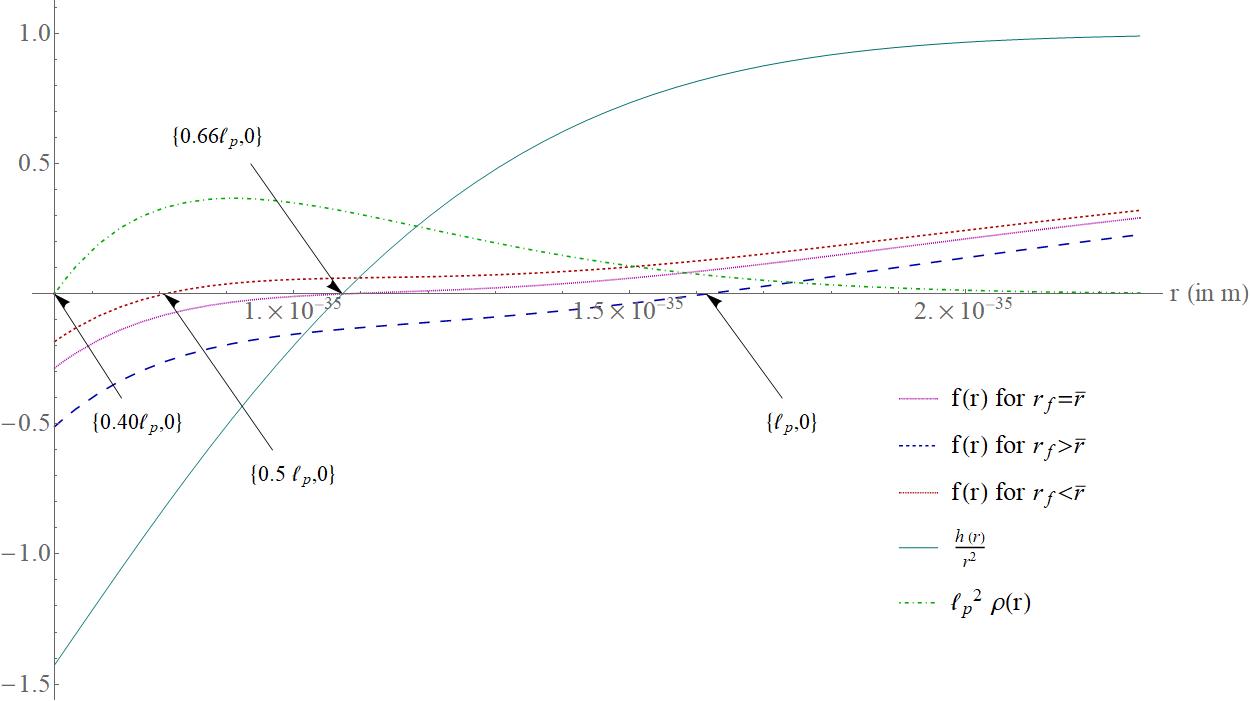}
\caption{Plot of $f(r)$, $\frac{h(r)}{r^2}$ and $l_p^2\rho(r)$ (eq.(\ref{1.44})) in the range $r_\rho<r<\frac{7l_p}{5}$. Here, $f(r)$ is plotted for three cases: 1) when $r_f=\bar{r}$ for which $M=\sqrt{\frac{9 \ln 2}{8 \pi}}m_p$, 2) $r_f=l_p>\bar{r}$ for which $M=\frac{m_p}{2}\left(1+2e^{-\pi}\right)$, and 3) $r_f=0.50 l_p<\bar{r}$ for which $M=\frac{m_p}{4}\left(1+2e^{-\frac{\pi}{4}}\right)$. It is evident from the plot that at $r=\bar{r}$, $f(\bar{r})=h(\bar{r})=0$. In the plot, the $x$ axis denotes the radial distance from the $r=0$ point and the $y$ axis denotes the amplitude of the dimensionless functions. The $\{r_\rho,0\}$ and $\{\bar{r},0\}$ points are marked separately in the plot.}\label{All_Plot_2}
\end{center}
\end{figure}

\noindent For $n=\frac{1}{2},1$ (in eq.(\ref{1.22})), we have
\begin{align}
	f(r)=&1-\frac{2GM}{rc^2}+\frac{ l_p}{\sqrt{\pi}r}e^{-\frac{\pi r^2}{l_p^2}}~,~~h(r)=r^2-\frac{2l_p r}{\sqrt{\pi}}e^{-\frac{\pi r^2}{l_p^2}}~,~~(n=1/2)\label{1.43}\\
	f(r)=&1-\frac{2GM}{rc^2}+2 e^{-\frac{\pi r^2}{l_p^2}}~,~~h(r)=r^2\left(1-4e^{-\frac{\pi r^2}{l_p^2}}\right)~,~~\left(n=1\right)\label{1.44}~.
\end{align}
Similar to the $n=0$ case, we can list the corresponding energy momentum tensors for the above two cases when $h(r)=r^2$. 
\noindent The matter density functions for the above two metric structures can be obtained from the Einstein field equations and read
\begin{align}
\rho(r)&=\frac{1}{4l_pr\sqrt{\pi}}e^{-\frac{\pi r^2}{l_p^2}}~,~~(n=1/2)\label{1.43a}\\
\rho(r)&=\frac{1}{2l_p^2}\left(1-\frac{l_p^2}{2\pi r^2}\right)e^{-\frac{\pi r^2}{l_p^2}}~,~~(n=1)\label{1.44a}~.
\end{align}
We start by discussing some important features regarding the metric functions in eq.(\ref{1.44}). We first note that the non-trivial value for which $h(r)=0$ is for $\bar{r}=l_p\sqrt{\frac{2\ln 2}{\pi}}\simeq 0.66 l_p$. From eq.(\ref{1.44a}), it is evident that the value of $r$ below which the energy density becomes negative is $r_\rho=\frac{l_p}{\sqrt{2\pi}}\simeq0.40l_p$. It is important to observe that $\bar{r}>r_\rho$ and as a result, for $r=\bar{r}$ which is less than $l_p$, the energy density remains positive. For the metric in eq.(\ref{1.44}), the condition $h(r)=0$ gives rise to three distinct cases, namely, the zero of $f$ is greater than $\bar{r}$, equals to $\bar{r}$, and less than $\bar{r}$. Note that for black holes with larger masses $r_f\gg \bar{r},r_\rho$. We first want to investigate the case, when the event horizon radius ($r_f$) is equal to $\bar{r}$. In this case, we obtain
\begin{equation}\label{wh1}
\begin{split}
f(\bar{r})&=1-\frac{2 G M}{\bar{r}c^2}+2e^{-\frac{\pi \bar{r}^2}{l_p^2}}=0\\
\implies M&=\frac{3l_pc^2}{4G}\sqrt{\frac{\ln 4}{\pi}}\\
&=\sqrt{\frac{9\ln 2}{8\pi}}m_p\\
&\simeq 0.50 m_p~.
\end{split}
\end{equation}
Hence, for a black hole of mass approximately equal to half of Planck's mass, the event horizon radius turns out to be the same as that of the radius for which $h(r)$ vanishes. We shall now look for the case, when $r_f$ is smaller than $\bar{r}$. It is although important to keep in mind that $r_f$ can never be smaller than $r_\rho$ as it will lead to negative energy density. We shall therefore  take $r_f=\frac{l_p}{2}$ so that $r_\rho<r_f<\bar{r}$. A straight forward calculation yields $M=m_p\left(\frac{1+2e^{-\frac{\pi}{4}}}{4}\right)\simeq 0.48 m_p$. It is important to note that the mass of the black hole is now smaller than the case when $r_f=\bar{r}$. This indicates to us that with the decreasing mass of the black hole the event horizon radius becomes smaller and even become smaller than the radius for which the $h(r)$ becomes zero. We shall now determine how small the mass of the black hole can be from our analysis. To determine this, we substitute $r=r_\rho$ in eq.(\ref{1.44}) and claim $f(r_\rho)=0$. The analysis gives $M_{\text{min}}=\frac{1}{2\sqrt{2\pi}}\left(1+2 e^{-\frac{1}{2}}\right)\simeq0.44 m_p$. The mass of the black hole cannot go below this value or a physical black hole will not exist anymore as the energy density in eq.(\ref{1.44a}) will become negative below this value. We plot $f(r)$, $\frac{h(r)}{r^2}$ and $l_p^2\rho(r)$ for $M=\sqrt{\frac{9 \ln2}{8 \pi}}m_p$ in the range $r_\rho<r<\frac{7 l_p}{5}$ in Fig. (\ref{All_Plot_2}) and observe the behaviour of the dimensionless functions. We also plot the other two cases $r_f>\bar{r}$ and $r_f<\bar{r}$ in Fig.(\ref{All_Plot_2}). For example, we have taken $r_f$ to be equal to the Planck's length $l_p$, this gives us the mass of the black hole to be equal to $M=\frac{m_p}{2}\left(1+2e^{-\pi}\right)$ and for the other case $r_f<\bar{r}$, the mass of the black hole comes out to be $M=\frac{m_p}{4}\left(1+e^{-\frac{\pi}{4}}\right)$ when $r_f=0.50l_p.$ The metric structure in eq.(\ref{1.44}) indicates the existence of Morris-Thorne type wormholes with the throat of the wormhole being located at $r=\bar{r}$ \cite{MorrisThorne} (for the case when $r_\rho<r_f<\bar{r}$). For the case when $r_f>\bar{r}$, the $r$ coordinate becomes timelike in nature inside the event horizon radius $r_f$. This may indicate towards a path which can take someone backwards in time. As this kind of behaviour is unphysical in nature, it is better not to associate this with any physical explanations.

\begin{figure}[ht!]
\begin{center}
\includegraphics[scale=0.38]{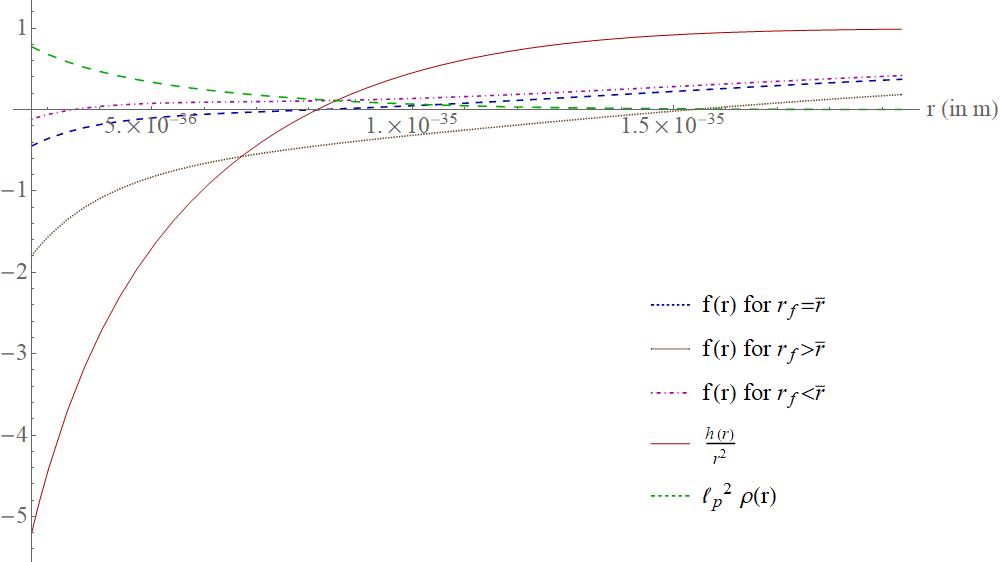}
\caption{Plot of $f(r)$, $\frac{h(r)}{r^2}$, and $l_p^2\rho(r)$  (eq.(\ref{1.43}))  in the range $\frac{l_p}{6}<r<\frac{6l_p}{5}$.  Here, $f(r)$ is plotted for three cases: 1) when $r_f=\bar{r}$ for which $M\simeq0.38m_p$, 2) $r_f>\bar{r}$ with $M\simeq0.49$, and 3) $r_f<\bar{r}$ with $M\simeq0.35m_p$. From the plot, we observe that at $r=\bar{r}\simeq 0.51l_p\simeq 8.16\times10^{-36}m$, $f(\bar{r})=h(\bar{r})=0$. In the plot, the $x$ axis denotes the radial distance from the $r=0$ point and the $y$ axis denotes the amplitude of the dimensionless functions.}\label{All_Plot_1}
\end{center}
\end{figure}

\noindent We now come back to the metric given by eq.(\ref{1.43}). This is much more straightforward as the energy density in eq.(\ref{1.43a}) corresponding to the lapse function $f(r)$ and $h(r)$ in eq.(\ref{1.43}) can never be negative for any value of $r$. A plot of $h(r)$ against $r$, reveals the value of $\bar{r}$ (where $h(r)=0$) to be approximately equal to $0.51 l_p$. Setting $f(\bar{r})=0$ gives the value of the mass of the black hole $M\approx0.38~m_p$. A plot of $f(r)$, $\frac{h(r)}{r^2}$, and $l_p^2\rho(r)$ for $M\approx0.38m_p$ is given in Fig.(\ref{All_Plot_1}). Here again, $f(r)$ is plotted for the three cases as has been done in Fig.(\ref{All_Plot_2}).

\noindent Eq.(s)(\ref{1.42},\ref{1.43},\ref{1.44}) have structures similar to that of a Schwarzschild metric along with a correction term consisting of an exponential factor multiplied by a coefficient. This coefficient consists of terms either proportional to some arbitrary powers of $\frac{r}{l_p}$ or some combination of these depending on the choice of the $n$ value.
Comparing eq.(s)(\ref{1.42},\ref{1.43},\ref{1.44}) with eq.(\ref{0.7a}), we observe striking similarities between the metric structures. It is evident that the exponential correction to the Bekenstein-Hawking black hole entropy leads to black hole geometries which have structures similar to that of a noncommutative inspired Schwarzschild black hole. It is also very important to observe that the matter density function obtained in eq.(\ref{1.42a},\ref{1.43a},\ref{1.44a}) has also striking similarities to that of the matter density function in \cite{Nicolini}. For the sake of completeness, we would like to mention that our approach leads to the noncommutative inspired Schwarzschild black hole in eq.(\ref{0.7a}) starting from eq.(\ref{0.8}), and also the appropriate matter density function getting fixed from the Einstein field equation. 

\begin{figure}[ht!]
\begin{center}
\includegraphics[scale=0.25]{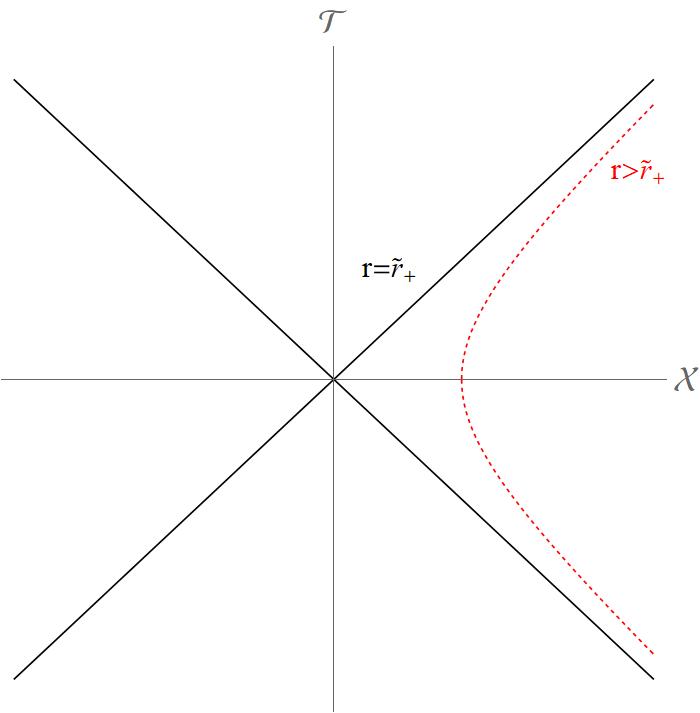}
\caption{$\mathcal{T}$ vs $\mathcal{X}$ plot: $\mathcal{T}=\pm\mathcal{X}$ lines correspond to the event horizon of the black hole at $r=\tilde{r}_+$  and the red dotted line correspond to a line of constant $r$ outside the event horizon of the black hole.}\label{Fig2}
\end{center}
\end{figure}

\noindent We shall now briefly investigate the conformal structure of the black hole $ds^2=-f(r)c^2dt^2+f(r)^{-1}dr^2$ in 1+1- spacetime dimensions for the form of $f(r)$ given in eq.(\ref{1.42}). For a near horizon expansion of the lapse function $f(r)$ about the modified horizon radius $\tilde{r}_+$ (given in eq.(\ref{1.25})), we can rewrite the line element as
\begin{equation}\label{1.45a}
ds^2=-(r-\tilde{r}_+)f'(\tilde{r}_+)c^2dt^2+\frac{1}{(r-\tilde{r}_+)f'(\tilde{r}_+)}dr^2~.
\end{equation}
With the help of the coordinate transformation $\rho=2\sqrt{\frac{r-\tilde{r}_+}{f'(\tilde{r}_+)}}$ we can recast the above line element in the Rindler form as
\begin{equation}\label{1.45b}
ds^2=-\frac{\rho^2{f'}^{2}(\tilde{r}_+)}{4}c^2dt^2+d\rho^2~.
\end{equation} 
We can now rewrite the above line element in the usual Minkowski spacetime in $1+1$-dimensions using the following coordinate transformations given by
\begin{align}
\mathcal{T}&=\frac{\rho}{c}\sinh\left(\frac{ctf'(\tilde{r}_+)}{2}\right)=\frac{\rho}{c}\sinh\left[\frac{ct}{2r_+}\left(1-e^{-\frac{\pi r_+^2}{l_p^2}}\right)\right]\label{1.45c}\\
\mathcal{X}&=\rho\cosh\left(\frac{ctf'(\tilde{r}_+)}{2}\right)=\rho\cosh\left[\frac{ct}{2r_+}\left(1-e^{-\frac{\pi r_+^2}{l_p^2}}\right)\right]\label{1.45d}~.
\end{align} 
In terms of the above coordinate transformations, we can rewrite the line element in eq.(\ref{1.45b}) as $ds^2=-c^2d\mathcal{T}^2+d\mathcal{X}^2$. It can be checked easily from Fig.(\ref{Fig2}) that for $r=\tilde{r}_+$, $\mathcal{T}=\pm\mathcal{X}$, the event horizon of the black hole is a null surface.

\section{Penrose-Carter diagram for the obtained metric function}
We shall now look at the causal structure of the spacetime geometry obtained in eq.(\ref{1.42}) and draw the corresponding Penrose diagram.
In 3+1-spacetime dimensions, the form of the line element corresponding to the lapse function $f(r)$ given is eq.(\ref{1.42}) is given by
\begin{equation}\label{conf1}
\begin{split}
ds^2=&-f(r)c^2dt^2+f(r)^{-1}dr^2+h(r)d\Omega^2~.
\end{split}
\end{equation}  
In order to obtain the Penrose diagram, we shall follow the analysis given in \cite{Graves,Arraut}. At first we consider a form of the metric given as 
\begin{equation}\label{conf2}
ds^2=-\kappa(u,v)(dv^2-du^2)+h(u,v)d\Omega^2
\end{equation}
where $\kappa(u,v)\neq0$.
Using $du=\frac{\partial u}{\partial t}dt+\frac{\partial u}{\partial r}dr$ and $dv=\frac{\partial v}{\partial t}dt+\frac{\partial v}{\partial r}dr$ in the above equation and comparing it with eq.(\ref{conf1}), we obtain the following set of equations
\begin{align}
(\partial_tv)^2-(\partial_t u)^2&=\frac{f(r)}{\kappa^2(u,v)}\label{conf3a}~,\\
(\partial_rv)^2-(\partial_r u)^2&=-\frac{1}{f(r)\kappa^2(u,v)}\label{conf3b}~,\\
\partial_t v\partial_r v-\partial_t u\partial_r u&=0\label{conf3c}~.
\end{align}
In terms of the Tortoise coordinate $r_*=\int\frac{dr}{f(r)}$ and with the redefinition $N(r_*)=\frac{f(r)}{\kappa^2}$ we can recast eq.(s)(\ref{conf3a}-\ref{conf3c}) as 
\begin{align}
(\partial_tv)^2-(\partial_t u)^2&=N(r_*)\label{conf4a}~,\\
(\partial_{r_*}v)^2-(\partial_{r_*} u)^2&=-N(r_*)\label{conf4b}~,\\
\partial_t v\partial_{r_*} v-\partial_t u\partial_{r_*} u&=0\label{conf4c}~.
\end{align}
Using eq.(s)(\ref{conf4a},\ref{conf4b},\ref{conf4c}), we arrive at the following two relations
\begin{align}
(\partial_t v+\partial_{r_*} v)^2&=(\partial_t u+\partial_{r_*} u)^2~,\label{conf5a}\\
(\partial_t v-\partial_{r_*} v)^2&=(\partial_t u-\partial_{r_*} u)^2\label{conf5b}~.
\end{align}
The determinant of the Jacobian for the change of coordinates from $\{t,r,\theta,\phi\}\rightarrow\{v,u,\theta,\phi\}$ attains a non-zero value for the two equations deduced from eq.(s)(\ref{conf5a},\ref{conf5b}) given below as
\begin{align}
\partial_t v+\partial_{r_*} v&=\partial_t u+\partial_{r_*} u~,\label{conf6a}\\
\partial_t v-\partial_{r_*} v&=-\partial_t u+\partial_{r_*} u\label{conf6b}~.
\end{align} 
From the above two relations one can obtain the two wave equations given by
\begin{equation}\label{conf7}
\partial_t^2 u(t,r_*)-\partial^2_{r_*}u(t,r_*)=0~,~\partial_t^2 v(t,r_*)-\partial^2_{r_*}v(t,r_*)=0~. 
\end{equation}
Eq.(\ref{conf7}) leads to a simple solution of $u(t,r_*)$ and $v(t,r_*)$ given by
\begin{align}
u(t,r_*)=\mathcal{A}(r_*+t)+\mathcal{B}(r_*-t)~,\label{conf8a}\\
v(t,r_*)=\mathcal{A}(r_*+t)-\mathcal{B}(r_*-t)~.\label{conf8b}
\end{align}
Substituting eq.(s)(\ref{conf8a},\ref{conf8b}) back in eq.(s)(\ref{conf4a}-\ref{conf4c}), we obtain the following relation
\begin{equation}
\begin{split}
2\frac{d}{dy_+}\left[\ln\left[\frac{d\mathcal{A}(y_+)}{dy_+}\right]\right]&=2\frac{d}{dy_-}\left[\ln\left[\frac{d\mathcal{B}(y_-)}{dy_-}\right]\right]\\&=\frac{d}{dr_*}\left(\ln N(r_*)\right)
\end{split}
\end{equation}
where $y_+=r_*+t$, $y_-=r_*-t$. Using the separation constant to be $2\beta$ in the above equation, we obtain the form of $\mathcal{A}(r_*+t)$, $\mathcal{B}(r_*+t)$, and $N(r_*)$ as follows
\begin{align}
\mathcal{A}(r_*+t)=&\frac{a_1e^{\beta(r_*+t)}}{\beta}+a_2~,\label{conf9a}\\
\mathcal{B}(r_*-t)=&\frac{a_3e^{\beta(r_*-t)}}{\beta}+a_4~,\label{conf9b}\\
N(r_*)=&a_5e^{2\beta r_*}\label{conf9c}~.
\end{align}
where $a_1,~a_2,~a_3,~a_4$ and $a_5$ are integration constants. Now a simple choice of these constants are $a_1=a_3=\frac{\beta}{2}$, and $a_2=a_4=0$ \cite{Arraut}. With this choice of the constants, we can obtain the forms of $u$, $v$, and $\kappa^2$  as
\begin{align}
u(t,r_*)&=e^{\beta r_*}\cosh \beta t~,~v(t,r_*)=e^{\beta r_*}\sinh \beta t~,\label{conf10a}\\
\kappa^2(r_*)&=\frac{f(r)}{\beta^2}e^{-2\beta r_*}~.\label{conf10b}
\end{align}
It is straightforward to understand that the radial null geodesics are described by $\frac{du}{dv}=\pm 1$.
In the regime $r>\tilde{r}_+$, it can be found that $\beta=\frac{f'(\tilde{r_+})}{2}$ \cite{Arraut} where the form of $\tilde{r}_+$  is given by eq.(\ref{1.25}). It is easy to show that in the limit $r\rightarrow\infty$, $u=\pm v$. One can now implement a set of new coordinate transformations, $\bar{p}=u+v~,~\bar{q}=u-v$ and eventually $T=\tan^{-1}\bar{p}~,~Z=\tan^{-1}\bar{q}$. The final choice of coordinate transformations are $\tilde{U}=T+Z$ and $\tilde{V}=T-Z$ using which we will draw the Penrose-Carter diagram for the black hole spacetime with the lapse function $f(r)$ given in eq.(\ref{1.42}). In this $\{\tilde{V},\tilde{U}\}$ coordinates, $\tilde{U}=\pm\tilde{V}$ denotes the event horizon $r=\tilde{r}_+$ for the metric function $f(r)$ given in eq.(\ref{1.42}). It is important to understand that the form of $f(r)$ given in eq.(\ref{1.42}) is valid only beyond the event horizon $r=\tilde{r}_+$ of the black hole. The future timelike infinity $i^+$ is described by $\{\tilde{V}=\frac{\pi}{2},\tilde{U}=\frac{\pi}{2}\}$ point and the past timelike infinity is given by $\{\tilde{V}=\frac{\pi}{2},\tilde{U}=-\frac{\pi}{2}\}$ point. The spatial infinity is given by $\{\tilde{V}=\pi,\tilde{U}=0\}$ point in the $\{\tilde{V},\tilde{U}\}$ coordinates. 

\begin{figure}[ht!]
\begin{center}
\includegraphics[scale=0.7]{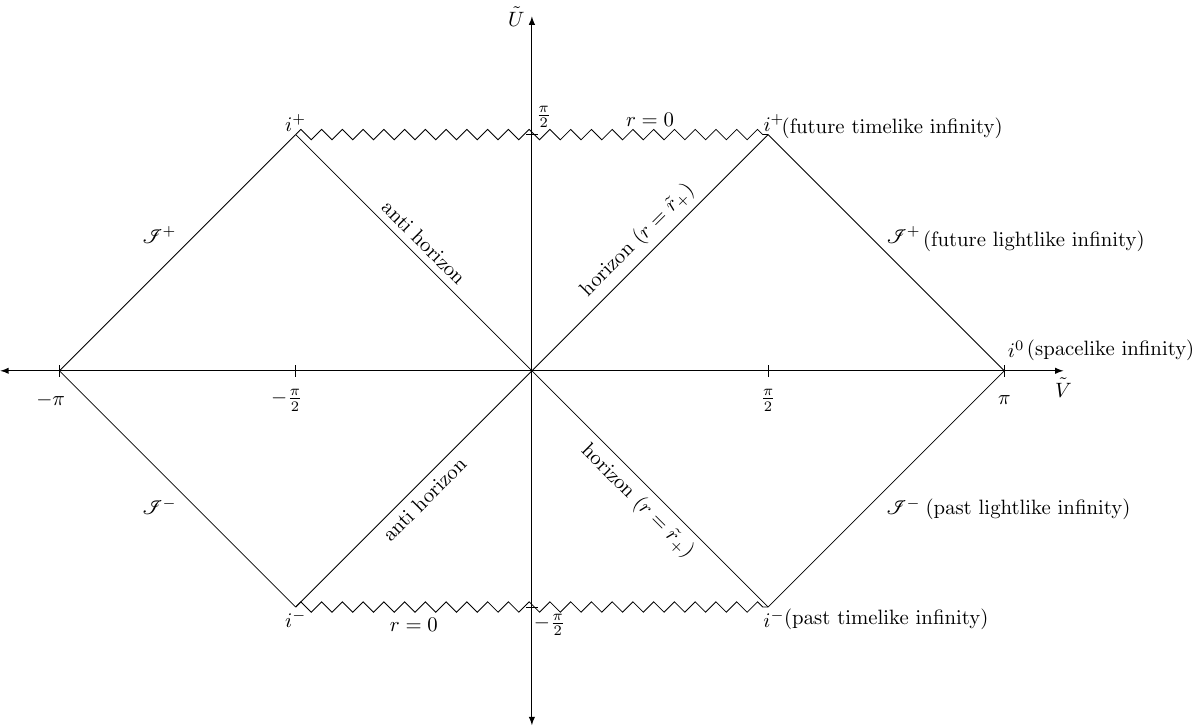}
\caption{Penrose-Carter diagram for the spacetime geometry in eq.(\ref{1.42})}\label{Fig3}
\end{center}
\end{figure}

\noindent We shall now investigate the singularity structure for the metric in eq.(\ref{1.42}). At first we need to obtain the form of the tortoise coordinate $r_*$ when $r\rightarrow 0$. Here, 
\begin{equation}
r_*=\int \left(1-\frac{r_+}{r}+\frac{l_p^2}{2\pi r^2}e^{-\frac{\pi r^2}{l_p^2}}\right)^{-1}dr=r+r_+\ln\left|\frac{r}{r_+}-1\right|-\frac{l_p^2}{2\pi}\int \frac{dr e^{-\frac{\pi r^2}{l_p^2}}}{(r-r_+)^2}+\text{constant}.
\end{equation}
In the limit $r\rightarrow 0$, the first two terms vanish whereas the third term can be computed as $\lim_{\epsilon\rightarrow 0}\int^\epsilon \frac{dr e^{-\frac{\pi r^2}{l_p^2}}}{(r-r_+)^2}\simeq\lim_{\epsilon\rightarrow 0}\frac{\sqrt{\pi}}{2r_+^2}\text{Erf}(\epsilon)=0$ (`Erf' denotes the Gauss error function).\footnote{In \cite{Ghosh}, the exponential corrections were shown to be dominant for microscopic black holes. In case of such a black hole, $r-\tilde{r}_+$ is a very small quantity for which one can generally use the near horizon expansion while computing $r_*$. One obtains $r_*=\frac{1}{f'(\tilde{r}_+)}\left(r+\ln\left|\frac{r}{\tilde{r}_+}-1\right|\right)$~. Hence, again $r_*$ goes to zero in the $r\rightarrow 0$ limit.} Hence, in the limit $r\rightarrow 0$, $r_*\rightarrow 0$. For $r_*=0$, $\tilde{U}=\tan^{-1}\left[e^{\beta t}\right]+\tan^{-1}\left[e^{-\beta t}\right]$, and $\tilde{V}=\tan^{-1}\left[e^{\beta t}\right]-\tan^{-1}\left[e^{-\beta t}\right]$. As $\tan^{-1}\theta+\tan^{-1}\frac{1}{\theta}=\frac{\pi}{2}$, we get that in the $r\rightarrow 0$ limit, $\tilde{U}=\frac{\pi}{2}$ whereas $\tilde{V}$ can vary based on the value of $t$ chosen. As $t\rightarrow 0$, $\tilde{V}=\frac{\pi}{2}$. This indicates that $r=0$ singularity is a straight line parallel to the $\tilde{V}$ axis extending from the $\left(0,\frac{\pi}{2}\right)$ to $\left(\frac{\pi}{2},\frac{\pi}{2}\right)$ point on the $(\tilde{V},\tilde{U})$ plane. One can now extend this to the left side. In this way, the Penrose-Carter diagram for the black hole spacetime with the lapse function $f(r)$ from eq.(\ref{1.42}) can be obtained and is given in Fig.(\ref{Fig3}).

\noindent Now we recall the Komar energy of a static, spherically symmetric black hole. It reads \cite{Komar,Wald,Smarr}
\begin{equation}\label{1.46}
E=\frac{1}{2}r^2\partial_r f(r)\biggr|_{r=\tilde{r}_+}=2ST
\end{equation} 
where $S$ and $T$ corresponds to the modified entropy and Hawking temperature of the black hole.

\noindent The Komar energy for the metric (given in eq.(\ref{1.42})) upto $\mathcal{O}(\theta)$ reads
\begin{equation}\label{1.47}
E=M\left[1-\left(1+\frac{\theta}{4\pi M^2}\right)e^{-\frac{4\pi M^2}{\theta}}\right]
\end{equation}
where we have replaced $l_p^2$ by $\theta$.
We now rewrite eq.(\ref{1.47}) in terms of the Bekenstein-Hawking entropy and Hawking temperature of 
the standard Schwarzschild black hole (upto $\mathcal{O}(\theta)$) as follows
\begin{equation}\label{1.48}
E=2 S_{Sch.} T_{Sch.}\left[1-\left(1+\frac{\theta}{S_{Sch.}}\right)e^{-\frac{S_{Sch.}}{\theta}}\right].
\end{equation}
This is the Smarr formula \cite{Smarr} for the modified black hole geometry and is another important result in our paper. Interestingly, the above formula has a similar form to the corresponding Smarr formula obtained in case of the noncommutative inspired Schwarzschild black hole \cite{SGRB}. 

\noindent Now we shall investigate some astrophysical aspects for one of the metric structures obtained. We consider the metric obtained in eq.(\ref{1.42})  and compute the effective potential and the photon radius for this metric. The Killing vector $K_\mu$ has the property $K_\mu u^{\mu}=\text{constant}$. For a massless particle, we have the relation $g_{\mu\nu}u^\mu u^\nu=0$. Using the time-like Killing vector $K_\mu=[-f(r),0,0,0]$ and the $\phi$-like Killing vector $L_\mu=[0,0,0,r^2\sin^2\theta]$ along with the condition $\theta=\frac{\pi}{2}$, one can obtain the effective potential for the lapse function in eq.(\ref{1.42}) to be
\begin{equation}\label{1.45}
V_{eff}=\frac{f(r)L^2}{r^2}-E^2=\frac{L^2}{r^2}\left[1-\frac{2GM}{rc^2}+\frac{l_p^2}{2\pi r^2}e^{-\frac{\pi r^2}{l_p^2}}\right]-E^2
\end{equation}
where $L$ is the total angular momentum and $E$ is the energy.
The radius of the photon sphere can be obtained from the conditions $\frac{dV_{eff}}{dr}\vert_{r=r_p}=0$, $\frac{d^2V_{eff}}{dr^2}\vert_{r=r_p}<0$, where  $r_p$ is the photon sphere radius. The photon sphere radius for the lapse function in  eq.(\ref{1.42}) reads
\begin{equation}\label{1.46a}
r_p=\frac{3GM}{c^2}\left[1-\frac{1}{2}\left(1+\frac{2c^4l_p^2}{9\pi G^2 M^2}\right)e^{-\frac{9\pi G^2M^2}{l_p^2c^4}}\right]~.
\end{equation}
The above result shows that there is a very small correction to the photon sphere radius coming from the exponential correction in the area law for the entropy of the black hole. 
\vskip 0.08cm
\section{Quantum corrected Einstein equations}\label{QEin}
While writing down the Einstein tensor's in eq.(s)(\ref{Ein00}-\ref{Ein33}), we have made use of the standard Einstein equations that can be obtained using the line element in eq.(\ref{1.23}). 
It is quite straightforward to understand that 
the new modified forms of $f(r)$ and $h(r)$ do not lead to vacuum solutions to the Einstein equations, rather they indicate towards a non-vanishing energy momentum tensor. Another way to incorporate such a metric structure from eq.(s)(\ref{1.42},\ref{1.422}) as a solution of vacuum Einstein field equations is to modify the Einstein equations. The easiest way is to subtract the right hand side of eq.(s)(\ref{Ein00}-\ref{Ein33}) from the left hand side and call them as quantum corrected Einstein equations. These equations read
\begin{align}
\bar{G}_{tt}=&f(r)\left[\frac{f(r)h^{'2}(r)}{4h^2(r)}-\frac{f(r)h''(r)}{h(r)}+\frac{1}{h(r)}-\frac{f'(r)h'(r)}{2h(r)}\right]-\frac{f(r)}{r^2}\left[\left(1-\frac{r_+}{r}\right)
\left[1+\frac{l_p^2}{\pi r^2}+\frac{4\pi r^2}{l_p^2}\right]+\frac{l_p^2}{2\pi r^2}\right]e^{-\frac{\pi r^2}{l_p^2}}=0~,\label{QEin00}\\
\bar{G}_{rr}=&\frac{1}{f(r)}\left[\frac{f(r)h^{'2}(r)}{4h^2(r)}-\frac{1}{h(r)}+\frac{f'(r)h'(r)}{2h(r)}\right]-\frac{1}{f(r)r^2}\left[\left(1-\frac{r_+}{r}\right)\left(1+\frac{l_p^2}{\pi r^2}\right)-\frac{l_p^2}{2\pi r^2}\right]e^{-\frac{\pi r^2}{l_p^2}}=0~,\label{QEin11}\\
\bar{G}_{\theta\theta}=&\frac{1}{4}\left[2f'(r)h'(r)+2f(r)h''(r)+2h(r)f''(r)-\frac{f(r)h^{'2}(r)}{h(r)}\right]\nonumber\\&+\frac{h(r)}{2r^2}\left[\left(1-\frac{r_+}{r}\right)\left(1+\frac{l_p^2}{\pi r^2}+\frac{2\pi r^2}{l_p^2}\right)+\frac{3r_+}{r}\left(1+\frac{l_p^2}{\pi r^2}+\frac{2\pi r^2}{3 l_p^2}\right)\right]e^{-\frac{\pi r^2}{l_p^2}}=0~,\label{QEin22}\\
\bar{G}_{\phi\phi}=&\frac{\sin^2\theta}{4}\left(2f'(r)h'(r)+2f(r)h''(r)+2h(r)f''(r)-\frac{f(r)h^{'2}(r)}{h(r)}\right)\nonumber\\&+\frac{h(r)\sin^2\theta}{2r^2}\left[\left(1-\frac{r_+}{r}\right)\left(1+\frac{l_p^2}{\pi r^2}+\frac{2\pi r^2}{l_p^2}\right)+\frac{3r_+}{r}\left(1+\frac{l_p^2}{\pi r^2}+\frac{2\pi r^2}{3 l_p^2}\right)\right]e^{-\frac{\pi r^2}{l_p^2}}=0~.\label{QEin33}
\end{align}
Eq.(s)(\ref{QEin00}-\ref{QEin33}) are our proposals for modified Einstein equations for which eq.(\ref{1.23}) serves as the vacuum solution with $f(r)$ and $h(r)$ being given by eq.(\ref{1.42}) and eq.(\ref{1.422}). In Appendix 1, we have listed the modified Einstein equations when $h(r)=r^2$. In Appendix 2, we have shown the connection between these modified Einstein equations with the first law of thermodynamics for the $h(r)=r^2$ case.
\section{Summary} 
In this work, we have formulated a way to obtain the metric structure of a static spherically symmetric black hole from the Bekenstein-Hawking black hole entropy with the event horizon radius of the black hole being given as an external input. To start with, we first applied our prescription to obtain the celebrated Reissner-Nordstr\"{o}m black hole metric from the \textit{``area by four"} law of black hole entropy and using the Einstein field equation, we have shown that the metric structure obtained in this procedure is indeed unique. The approach gets the solution from thermodynamics of black holes without solving the field equations. The field equations reveal that the solution must be unique for an appropriate energy-momentum tensor. This indicates that thermodynamics and gravity are related, a fact that have been discussed earlier in the literature \cite{TJacobson}. With this background, we then focussed on the modified Bekenstein-Hawking black hole entropy obtained from microstate counting of horizon states \cite{Ghosh}. Our method depends crucially on the input of the event horizon radius of the black hole.
Remarkably, the metric structures we obtain have striking similarities to that of a noncommutative inspired Schwarzschild black hole. Therefore, this metric would be consistent with a noncommutative inspired structure of spacetime. 
However, the geometry we obtain is not exactly identical to the noncommutative inspired Schwarzschild black hole \cite{Nicolini,Nicolini2,Chaichian,Nicolini3,SGRB}. 
We also obtain the underlying matter density function for the metric structure found in this procedure. It is also very important to note that the underlying matter density function also have striking structural similarities to that of the matter density function used in case of the noncommutative inspired Schwarzschild black holes \cite{Nicolini}. 
We also briefly discuss the conformal structure of the obtained spacetime geometry and draw the Penrose-Carter diagram for the spacetime geometry in eq.(\ref{1.42}). Next we compute the Komar energy for our obtained metric and from there we obtain the modified Smarr formula which corresponds to the modified Bekenstein-Hawking entropy. Once again this modified Smarr formula have resemblance to that of a noncommutative inspired Schwarzschild black hole \cite{SGRB}. We then discuss the astrophysical implications of our result by calculating the photon radius for such a geometry. We observe that there is a shift in the photon sphere radius due to exponential correction in the black hole entropy. We finally conclude our investigation by providing a set of vacuum Einstein field equations with quantum corrections which lead to one of the metrics obtained earlier from the modified entropy formula. In this case,  we consider that the form of the metric comes from the quantum corrected Einstein field equations without the existence of any matter distribution function. Finally, in Appendix 2, we show that the quantum corrected Einstein field equations can in principle be related to the first law of black hole thermodynamics which in principle establishes a nice connection between geometry and thermodynamics \cite{TPadmanabhan} (for $h(r)=r^2$ case).
\section*{Data availability statement}
The manuscript has no data associated with it.
\section*{Conflict of interest}
The authors declare no conflicts of interest.
\section*{Acknowledgment}
We thank the referees for making useful and valuable comments which have helped us to improve the manuscript substantially.
\section*{Apppendix 1: Quantum corrected Einstein equations for $h(r)=r^2$}
In the earlier part of our analysis after obtaining eq.(\ref{1.42}), we have considered the classical Einstein equation with matter density function $\rho(r)$ given in eq.(\ref{1.42a}) to prove the uniqueness of our solution. In this section, we look for quantum corrections to the Einstein equations. We now propose a set of modified Einstein vacuum field equations. The quantum Einstein vacuum field equations involving the $\{t,t\}$ and $\{r,r\}$ components of the quantum modified Einstein tensor  $\bar{G}_{tt}$ and $\bar{G}_{rr}$ are proposed as follows
\begin{equation}\label{1.49}
\begin{split}
r^2f(r)\bar{G}_{tt}=-r^2f(r)\bar{G}_{rr}=0
\end{split}
\end{equation}
where 
\begin{equation}\label{1.49a}
r^2f(r)\bar{G}_{tt}=\left[1-f(r)-rf'(r)\right]-\left(1+\frac{l_p^2}{2\pi r^2}\right)e^{-\frac{\pi r^2}{l_p^2}}~.
\end{equation}
Similarly, the other two quantum modified vacuum field equations are proposed as
\begin{equation}\label{1.50}
\begin{split}
\bar{G}_{\theta\theta}=-\frac{1}{\sin^2\theta}\bar{G}_{\phi\phi}=0
\end{split}
\end{equation}
where
\begin{equation}\label{1.50a}
\bar{G}_{\theta\theta}=\frac{r}{2}(2f'(r)+rf''(r))-\left(\frac{\pi r^2}{l_p^2}+\frac{1}{2}+\frac{l_p^2}{2\pi r^2}\right)e^{-\frac{\pi r^2}{l_p^2}}~.
\end{equation}
In eq.(s)(\ref{1.49},\ref{1.50}), $\bar{G}_{tt}$, $\bar{G}_{rr}$, $\bar{G}_{\theta\theta}$, and $\bar{G}_{\phi\phi}$ denotes the quantum modified Einstein tensor. Solving eq.(\ref{1.49}) or eq.(\ref{1.50}) perturbatively, we obtain the form of the metric $f(r)$ given in eq.(\ref{1.42}). It is important to observe that both eq.(s)(\ref{1.49},\ref{1.50}) are vacuum field equations.

\noindent It is important to note that there are no solid justifications for the quantum Einstein equations in eq.(\ref{1.49}-\ref{1.50a}) at this moment. In order to obtain the quantum corrections, we have absorbed the energy-momentum tensor into the left hand side of the Einstein field equations and proposed the resultant object as the quantum vacuum Einstein field equation giving rise to the solution $f(r)$ in eq.(\ref{1.42}).
\vskip 0.2cm
\section*{Appendix 2: Quantum corrected Einstein equations and its connection to the first law of black hole thermodynamics for $h(r)=r^2$} We shall now exploit the analytical form of $r^2f(r)\bar{G}_{rr}$ given in eq.(\ref{1.49}) around the horizon radius $\tilde{r}_+$ given in eq.(\ref{1.25}). In the limit $r\rightarrow\tilde{r}_+$, eq.(\ref{1.49}) can be expressed  (upto $\mathcal{O}\left(\exp\left(-\pi r_+^2/l_p^2\right)\right)$) as follows
\begin{equation}\label{1.51}
\tilde{r}_+f'(\tilde{r}_+)-1+\left(1+\frac{l_p^2}{2\pi r_+^2}\right)e^{-\frac{\pi r_+^2}{l_p^2}}=0~.
\end{equation}
We now multiply both sides of the above equation with  $\frac{c^4}{2G}d\tilde{r}_+$ and obtain the following relation
\begin{align}
\frac{c^4}{2G}f'(\tilde{r}_+)\tilde{r}_+d\tilde{r}_+&\simeq \frac{c^4}{2G}\left[d\tilde{r}_+-\left(1+\frac{l_p^2}{2\pi r_+^2}\right)e^{-\frac{\pi r_+^2}{l_p^2}}dr_+\right]\nonumber\\
&=\frac{c^4}{2G}dr_+\label{1.52}\\&=c^2dM~.\nonumber
\end{align}
The left hand side of the above equation can be rearranged as 
\begin{equation}\label{1.52a}
\frac{c^4}{2G}f'(\tilde{r}_+)\tilde{r}_+d\tilde{r}_+=\left(\frac{\hbar cf'(\tilde{r}_+)}{4\pi k_B}\right)\frac{k_Bc^3}{4\hbar G}d(4\pi\tilde{r}_+^2)~.
\end{equation}
 In this relation the term in the parenthesis is the temperature of the black hole whose metric is given by eq.(\ref{1.42}). Invoking the form of the horizon from eq.(\ref{1.25}), we finally obtain the following relation
\begin{equation}\label{1.53}
Td\left(\frac{k_BA}{4l_p^2}-k_Be^{-\frac{A}{4l_p^2}}\right)=c^2dM
\end{equation}
where we have identified $A=4\pi r_+^2$ with $r_+$ being the usual Schwarzschild radius. Now from eq.(\ref{1.22}), we can see that for $n=0$ case, the modified entropy formula is $S=\frac{k_BA}{4l_p^2}-k_Be^{-\frac{A}{4l_p^2}}$. With this identification, we can recast eq.(\ref{1.53}) given as
\begin{equation}\label{1.54}
TdS=c^2dM=dE
\end{equation}
which is the usual first law of black hole thermodynamics. The above analysis establishes a direct connection between geometry and thermodynamics \cite{TPadmanabhan}. It is also important to identify that the pressure term at the horizon is zero leading to the conclusion that we are dealing with a vacuum field equation. This indeed is a nice consistency check of the proposed quantum Einstein equations in eq.(s)(\ref{1.49},\ref{1.50a}). 
\vskip 0.08cm



\begin{thebibliography}{8}
	\bibitem{Einstein15}
	A. Einstein, ``\textit{Die Feldgleichungen der Gravitation}", Sitzungsber Preuss Akad Wiss (1915) 844.
	\bibitem{Einstein16}
	A. Einstein, ``\textit{Die Grundlage der allgemeinen Relativit\"{a}tstheorie}", \href{https://doi.org/10.1002/andp.19163540702}{Ann. der Physik 49 (1916) 769}.
	\bibitem{Bekenstein}
	J.D. Bekenstein, ``\textit{Black holes and the second law}", \href{https://doi.org/10.1007/BF02757029}{Lett. Nuovo Cimento 4 (1972) 737}.
	\bibitem{Bekenstein2}
	J. D. Bekenstein, ``\textit{Black Holes and Entropy}", \href{https://link.aps.org/doi/10.1103/PhysRevD.7.2333}{Phys. Rev. D 7 (1973) 2333}.
	\bibitem{Hawking}
	S.W. Hawking, ``\textit{Black hole explosions?}", \href{https://doi.org/10.1038/248030a0}{Nature 248 (1974) 30}.
	\bibitem{Hawking2}
	S.W. Hawking, ``\textit{Particle creation by black holes}", \href{https://doi.org/10.1007/BF02345020}{Commun. Math. Phys 43 (1975) 199}.
	\bibitem{Hawking3}
	S.W. Hawking, ``\textit{Black holes and thermodynamics}", \href{https://link.aps.org/doi/10.1103/PhysRevD.13.191}{Phys. Rev. D 13 (1976) 191}.
	\bibitem{Hawking4}
J. M. Bardeen, B. Carter and S. W. Hawking, ``\textit{The four laws of black hole mechanics}", \href{https://doi.org/10.1007/BF01645742}{Commun. Math. Phys. 31 (1973) 161}.
	\bibitem{ACV}
	D. Amati, M. Ciafaloni and G. Veneziano, ``\textit{Can spacetime be probed below the string size?}", \href{https://doi.org/10.1016/0370-2693(89)91366-X}{Phys. Lett. B.216 (1989) 41}.
	\bibitem{KOPAPR}
	K. Konishi, G. Paffuti and P. Provero, ``\textit{Minimum physical length and the generalized uncertainty principle in string theory}", \href{https://doi.org/10.1016/0370-2693(90)91927-4}{Phys. Lett. B 234 (1990) 276}.
		\bibitem{ROVELLI}
	C. Rovelli, ``\textit{Loop Quantum Gravity}", \href{https://doi.org/10.12942/lrr-1998-1}{Living. Rev. Relativ. 1 (1998) 1}.
	\bibitem{CARLIP}
	S. Carlip, ``\textit{Quantum gravity: a progress report}", \href{https://iopscience.iop.org/article/10.1088/0034-4885/64/8/301}{Rep. Prog. Phys. 64 (2001) 885}.
	\bibitem{Girelli}
	F. Girelli, E. R. Livine and D. Oriti, ``\textit{Deformed special relativity as an effective flat limit of quantum gravity}",  \href{https://doi.org/10.1016/j.nuclphysb.2004.11.026}{Nucl. Phys. B 708 (2005) 411}.
	\bibitem{Strominger}
	A. Strominger and C. Vafa, ``\textit{Microscopic origin of the Bekenstein-Hawking entropy}", \href{https://www.sciencedirect.com/science/article/pii/0370269396003450}{Phys. Lett. B 379 [(1996) 99}. 
	\bibitem{RD}
	R. Dijkgraaf, E. P. Verlinde and H. L. Verlinde, ``\textit{Counting dyons in N = 4 string theory}", \href{https://www.sciencedirect.com/science/article/pii/S0550321396006402}{Nucl. Phys. B 484 (1997) 543}.
	\bibitem{Ashtekar}
	A. Ashtekar, J. Baez, A. Corichi and K. Krasnov, ``\textit{Quantum Geometry and Black Hole Entropy}", \href{https://link.aps.org/doi/10.1103/PhysRevLett.80.904}{Phys. Rev. Lett. 80 (1998) 904}.
	\bibitem{ParthaKaul}
	R. K. Kaul and P. Majumdar, ``\textit{Logarithmic Correction to the Bekenstein-Hawking Entropy}", \href{https://link.aps.org/doi/10.1103/PhysRevLett.84.5255}{Phys. Rev. Lett. 84 (2000) 5255}.
	\bibitem{Lewandowski}
	M. Domagala and J. Lewandowski, ``\textit{Black-hole entropy from quantum geometry}", \href{http://dx.doi.org/10.1088/0264-9381/21/22/014}{Class. Quant. Gravit. 21 (2004) 5233}.
	\bibitem{Meissner}
	K. A. Meissner, ``\textit{Black-hole entropy in loop quantum gravity
}", \href{http://dx.doi.org/10.1088/0264-9381/21/22/015}{Class. Quant. Gravit. 21 (2004) 5245}.
	\bibitem{Ghosh2}
	A. Ghosh and P. Mitra, ``\textit{An improved estimate of black hole entropy in the quantum geometry approach}", \href{https://www.sciencedirect.com/science/article/pii/S0370269305006076}{Phys. Lett. B 616 (2005) 114}.
	\bibitem{MandalSen}
	I. Mandal and A. Sen, ``\textit{Black hole microstate counting and its macroscopic counterpart*}", \href{https://doi.org/10.1088/0264-9381/27/21/214003}{Class. Quant. Gravit. 27 (2010) 214003}.
	\bibitem{RBSGSKM}
	R. Banerjee, S, Gangopadhyay and S. K. Modak, ``\textit{Voros product, noncommutative Schwarzschild black hole and corrected area law}", \href{https://www.sciencedirect.com/science/article/pii/S037026931000208X}{Phys. Lett. B 686 (2010) 181}.
	\bibitem{Holkar}
	A. Dabholkar, J. Gomes and S. Murthy, ``\textit{Counting all dyons in $\mathcal{N}$=4
 string theory}",  \href{https://doi.org/10.1007/JHEP05(2011)059}{J. High Energy Phys. 2011 (2011) 59}. 
	\bibitem{SGDKR}
	S. Gangopadhyay and D. Roychowdhury, ``\textit{CORRECTED AREA LAW AND KOMAR ENERGY FOR NONCOMMUTATIVE INSPIRED REISSNER–NORDSTR\"{O}M BLACK HOLE}",  \href{https://doi.org/10.1142/S0217751X12500418}{Int. J. Mod. Phys. A 27 (2012) 1250041}.
	\bibitem{SG0}
	S. Gangopadhyay, ``\textit{VOROS PRODUCT AND NONCOMMUTATIVE INSPIRED BLACK HOLES}", \href{https://doi.org/10.1142/S0217732313500302}{Mod. Phys. Lett. A 28 (2013) 1350030}.
	\bibitem{Holkar2}
	A. Dabholkar, J, Gomes and S. Murthy, ``\textit{Nonperturbative black hole entropy and Kloosterman sums}",  \href{https://doi.org/10.1007/JHEP03(2015)074}{J. High Energy Phys. 2015 (2015) 74}.
\bibitem{Ghosh}
	A. Chatterjee and A. Ghosh, ``\textit{Nonperturbative black hole entropy and Kloosterman sums}", \href{https://link.aps.org/doi/10.1103/PhysRevLett.125.041302}{Phys. Rev. Lett. 125 (2020) 041302}.
\bibitem{Rovelli}
C. Rovelli, \textit{Quantum Gravity} (\textit{Cambridge Monographs on Mathematical Physics}), Cambridge University Press; Cambridge.
\bibitem{Doplicher}
S. Doplicher, K. Fredenhagen, and J. E. Roberts, ``\textit{Spacetime quantization induced by classical gravity}", \href{https://doi.org/10.1016/0370-2693(94)90940-7}{Phys. Lett. B 331 (1994) 39}.
\bibitem{Doplicher2}
S. Doplicher, K. Fredenhagen, and J. E. Roberts, ``\textit{The quantum structure of spacetime at the Planck scale and quantum fields}", \href{https://doi.org/10.1007/BF02104515}{Commun. Math. Phys. 172 (1995) 187}.
\bibitem{Douglas}
M. R. Douglas and N. A. Nekrasov, ``\textit{Noncommutative field theory}",  \href{https://link.aps.org/doi/10.1103/RevModPhys.73.977}{Rev. Mod. Phys. 73 (2001) 977}.
	\bibitem{Nicolini}
	P. Nicolini, A. Smailagic and E. Spallucci, ``\textit{Noncommutative geometry inspired Schwarzschild black hole}",  \href{https://www.sciencedirect.com/science/article/pii/S0370269305016126}{Phys. Lett. B 632 (2006) 547}.	
	\bibitem{Nicolini2}
	S. Ansoldi, P. Nicolini, A. Smailagic and E. Spalucci, ``\textit{Non-commutative geometry inspired charged black holes}", \href{https://www.sciencedirect.com/science/article/pii/S0370269306015607}{Phys. Lett. B 645 (2007) 261}.
	\bibitem{Chaichian}
	M. Chaichian, A. Tureanu and G. Zet, ``\textit{Corrections to Schwarzschild solution in noncommutative gauge theory of gravity}", \href{https://www.sciencedirect.com/science/article/pii/S037026930800097X}{Phys. Lett. B 660 (2008) 573}.
	\bibitem{Nicolini3}
	P. Nicolini, ``\textit{NONCOMMUTATIVE BLACK HOLES, THE FINAL APPEAL TO QUANTUM GRAVITY: A REVIEW}", \href{https://doi.org/10.1142/S0217751X09043353}{Int. J. Mod. Phys. A 24 (2009) 1229}.
	\bibitem{SGRB}
	R. Banerjee and S. Gangopadhyay, ``\textit{Komar energy and Smarr formula for noncommutative inspired Schwarzschild black hole}", \href{https://doi.org/10.1007/s10714-011-1250-2}{Gen. Relativ. Gravit. 43 (2011) 3201}.
\bibitem{Voros}
A. Voros, ``Wentzel-Kramers-Brillouin method in the Bargmann representation", \href{https://link.aps.org/doi/10.1103/PhysRevA.40.6814}{Phys. Rev. A 40 (1989) 6814}.
	\bibitem{Gradshteyn}
I. S. Gradshteyn and I. M. Ryzhik, \textit{Table of Integrals, Series and Products}, 7th edition (Academic Press, Amsterdam, 2007).
\bibitem{Arraut}
I. Arraut, D. Batic, and M. Nowakowski, ``\textit{Maximal extension of the Schwarzschild space-time inspired by noncommutative geometry}", \href{https://doi.org/10.1063/1.3317913}{J. Math. Phys. 51 (2010) 022503}.
\bibitem{TJacobson}
T. Jacobson, ``\textit{Thermodynamics of Spacetime: The Einstein Equation of State}", \href{https://link.aps.org/doi/10.1103/PhysRevLett.75.1260}{Phys. Rev. Lett. 75 (1995) 1260}.
\bibitem{Arbey}
A. Arbey, J. Auffinger, M. Geiller, E. R. Livine, and F. Sartini, ``\textit{Hawking radiation by spherically-symmetric static black holes for all spins:  Teukolsky equations and potentials}", \href{https://doi.org/10.1103/PhysRevD.103.104010}{Phys. Rev. D 103 (2021) 104010}.
\bibitem{MorrisThorne}
M. S. Morris and K. S. Thorne, ``\textit{Wormholes in spacetime and their use for interstellar travel: A tool for teaching general relativity}", \href{https://doi.org/10.1119/1.15620}{Am. J. Phys. 56 (1988) 395}.
\bibitem{Graves}
J. C. Graves and D. R. Brill, ``\textit{Oscillatory Character of Reissner-Nordström Metric for an Ideal Charged Wormhole}", \href{https://link.aps.org/doi/10.1103/PhysRev.120.1507}{Phys. Rev. 120 (1960) 1507}.
\bibitem{Komar}
A. Komar, ``\textit{Covariant Conservation Laws in General Relativity}", \href{https://link.aps.org/doi/10.1103/PhysRev.113.934}{Phys. Rev. 113 (1959) 934}.
\bibitem{Wald}
R. M. Wald, \textit{General Relativity}, University Press, Chicago, USA (1984).
\bibitem{Smarr}
L. Smarr, ``Mass Formula for Kerr Black Holes", \href{https://link.aps.org/doi/10.1103/PhysRevLett.30.71}{Phys. Rev. Lett. 30 (1973) 71}; Erratum, \href{https://link.aps.org/doi/10.1103/PhysRevLett.30.521}{Phys. Rev. Lett. 30 (1973) 521}.
\bibitem{TPadmanabhan}
T. Padmanabhan, ``\textit{Gravitation: Foundations and Frontiers}", Cambridge University Press, Cambridge (2010).
\end{thebibliography}
\end{document}